\documentclass[twocolumn,prb,showpacs,preprintnumbers,amsmath,amssymb]{revtex4}
\usepackage{amsmath,amsfonts,amssymb,graphics,graphicx,epsfig,color,times,indentfirst,layout,hyperref,subfigure}

\newcommand{\be}{\begin{equation}}
\newcommand{\ee}{\end{equation}}
\newcommand{\bma}{\begin{pmatrix}}
\newcommand{\ema}{\end{pmatrix}}
\newcommand{\balig}{\begin{align}}
\newcommand{\ealig}{\end{align}}

\newcommand{\ba}{\begin{eqnarray}}
\newcommand{\ea}{\end{eqnarray}}

\newcommand{\ignore}[1]{}

\hypersetup{linktocpage}

\begin{document}

\title{Unified model for conductance through DNA with the Landauer-B$\ddot{\textbf{u}}$ttiker formalism}
\date{\today}

\author{Jianqing Qi, Neranjan Edirisinghe, M. Golam Rabbani, and M. P. Anantram }
\affiliation{Department of Electrical Engineering, University of Washington, Seattle, WA 98195-2500}

\begin{abstract}
In this work, we model the zero-bias conductance for the four different DNA strands that were used in conductance measurement experiment [A. K. Mahapatro, K. J. Jeong, G. U. Lee and D. B. Janes, Nanotechnology, $\textbf{18}$, 195202 (2007)]. Our approach consists of three elements: (i) $ab initio$ calculations of DNA, (ii) Green's function approach for transport calculations and (iii) the use of two parameters to determine the decoherence rates. We first study the role of the backbone. We find that the backbone can alter the coherent transmission significantly at some energy points by interacting with the bases, though the overall shape of the transmission stays similar for the two cases. More importantly, we find that the coherent electrical conductance is tremendously smaller than what the experiments measure. We consider DNA strands under a variety of different experimental conditions and show that even in the most ideal cases, the calculated coherent conductance is much smaller than the experimental conductance. To understand the reasons for this, we carefully look at the effect of decoherence. By including decoherence, we show that our model can rationalize the measured conductance of the four strands, both qualitatively and quantitatively. We find that the effect of decoherence on $G:C$ base pairs is crucial in getting agreement with the experiments. However, the decoherence on $G:C$ base pairs alone does not explain the experimental conductance in strands containing a number of $A:T$ base pairs. Including decoherence on $A:T$ base pairs is also essential.  By fitting the experimental trends and magnitudes in the conductance of the four different DNA molecules, we estimate for the first time that the deocherence rate is 6 meV for $G:C$ and 1.5 meV for $A:T$ base pairs.  
\end{abstract}
\maketitle

\section{Introduction}

Molecular electronic devices have been increasingly attracting attention from researchers working on both top down and bottom up approaches. \cite{Zhitenev, Chen, Kim} Among various molecular candidates, DNA has the attractive features of recognition and self-assembly. \cite{Porath, Tu, Douglas} Many efforts \cite{Jortner, Giese1, Giese2, Barnett, CLiu, Lewis, Senthilkumar, Genereux, Slinker, Topics} have been devoted to understanding the mechanism of charge transfer and transport in DNA. From an electrical device perspective, double barrier resonant tunneling structures, \cite{Adessi} spin specific electron conductor, \cite{Zacharias, Naaman} field effect transistor, \cite{Yokoyama} trinary logic, \cite{Chakraborty} optomechanical molecular motor, \cite{McCullagh} negative differential resistance, \cite{Kang} detection of lesions by repair proteins \cite{Sontz} and doping \cite{Roh} are in principle possible with DNA.  Furthermore, recent work has also shown that it is possible to detect diseases by measuring the electrical conductivity of DNA. \cite{Tsutsui, Hihath} Researchers have also found that DNA can either be an insulator, \cite{Pablo} a semiconductor, \cite{Porathnature, Takagi} an ohmic conductor, \cite{Fink, Thong} or a superconductor, \cite{Kasumov, Bouchiat} depending on a variety of conditions. 
 
It is broadly agreed upon that while the understanding of the electrical conductance of DNA has matured over the last two decades, explaining experiments or making predictions  remains a challenge, compared to nanoengineered materials such as nanotubes and nanowires that have comparable dimensions. Modeling of the conductivity of a DNA molecule in solution is complex because many factors contribute to the charge transport process. In wet DNA, water molecules are critical in influencing the molecular structure and hence the electronic properties. A-DNA usually has five to ten water molecules per base, while for B-DNA more than thirteen water molecules per base are preferred. \cite{Endres} At different hydration levels, different local densities of states (DOS) are obtained. \cite{Barnett} Besides, because the backbone of a DNA molecule is negatively charged in its phosphate groups, it is believed that the molecule is surrounded by cations such as $Na^{+}$, $K^{+}$, $Mg^{2+}$ and $H_{3}O^{+}$. Previous studies have shown that variations in cation position \cite{Barnett} and type \cite{Adessiapl} can modulate the highest occupied molecular orbital (HOMO) and lowest unoccupied molecular orbital (LUMO) levels. Configuration changes (lattice vibrations/deformation) can also modify the conductivity of DNA molecules. Theoretical calculations have confirmed that changes in the distance and the angle between consecutive bases can cause variability in conduction channels by influencing the hybridization between the $\pi$ orbitals of adjacent bases. \cite{Adessi}  Additionally, the length and sequence play an important role in determining the conductivity of DNA.  The sequence is important \cite{EMeggers, KYoo} because of the different ionization potential of the four bases, adenine ($A$), thymine ($T$), guanine ($G$) and cytosine ($C$) and sequence dependent conduction. \cite{BXu, Mahapatro}

The conductivity of DNA molecules lying between two metal contacts has been studied by a few different groups. References \onlinecite{Adessi, Adessiapl, HMehrez, SSairam, Macia, RGutierrez, RGutierrez2, Cuniberti, VMalyshev, VMalyshev2, XWang, JYi} have examined the conductance of DNA molecules whose coordinates are frozen based on the Landauer-B$\ddot{\textup{u}}$ttiker formalism. Their approaches can be classified into two categories, the first of which uses density functional theory (DFT). \cite{Adessi, Adessiapl, HMehrez, SSairam} Because DFT calculation for long DNA strands is very time-consuming, researchers alternatively first carry DFT calculations on short strands (typically two to five base pairs), and extract the Hamiltonian and overlap matrices for the bases and their interactions with neighbors. One can then construct a larger matrix corresponding to the Hamiltonian for a longer strand from the sub-Hamiltonians obtained from the mentioned calculations on shorter strands. Another category is to use more simplified Hamiltonians that account for only one or two energy levels on each base and their interactions with energy levels in neighboring bases.\cite{Macia, RGutierrez, RGutierrez2, Cuniberti, VMalyshev, VMalyshev2, XWang, JYi}

The modeling of conductance in wet-DNA lying between metal contacts is a more difficult problem because the surrounding environment (water molecules, ions and conformation) fluctuates with time. In this case recent work by Ref. \onlinecite{Benjamin}  uses a combination of QM/MM to evaluate the role of fluctuating environment on DNA transport. These calculations freeze the location of the atoms at each sampling point in the MD simulation and calculate the phase-coherent conductance using the Landauer-Buttiker approach. In addition to the above mentioned work that uses the Landauer-Buttiker approach, there has also been work that model DNA using a rate equation involving hopping and tunneling depending on the sequence involved. \cite{Jortner, AVoityuk, Senthilkumar, MBixon, Voityuk}  

In this work, we study the conductance of dry DNA by carefully investigating the role of decoherence. Decoherence in dry DNA molecule mainly arises from time-dependent fluctuations where the electrons will lose phase information to the environment. Electrons in dry DNA can lose phase information by interacting with lattice vibrations and ambient electromagnetic fields. In this paper, we model the decoherence using the phenomenological B$\ddot{\textup{u}}$ttiker probes. \cite{Buttikerprl, Buttikeribm} We do this in the context of four different dry DNA strands considered in the experiments of Ref. \onlinecite{Mahapatro} and shown in Fig. \ref{Sequences}. The four strands, which will be further discussed in section II, contain zero, one, three and five $A:T$ base pairs in the middle part of a strand that otherwise consists of only $G:C$ base pairs. We note that the method of including decoherence is not unique. Four perceptive recent papers include the effect of decoherence in wet-DNA using completely different approaches. \cite{Kubar1, Zilly, Cuniberti, RGutierrez, RGutierrez2} Reference \onlinecite{Kubar1} represents the influence of atomic charges from the DNA backbone, water molecules and counterions with an empirical force field using a hybrid quantum mechanics-molecular mechanics (QM/MM) framework. Reference \onlinecite{Zilly} uses a new statistical decoherence model developed by their group. References \onlinecite{RGutierrez} and \onlinecite{RGutierrez2} use the harmonic phonon bath to describe the dissipative environment. Reference \onlinecite{Cuniberti} includes the effect of energetic vibronic coupling via a full-fledged nonequilibrium Green's function approach. In our approach the DNA strands are fixed and charge transport in DNA is a decoherent process. A self-energy is used to represent the decoherence. A similar approach has been adopted by Ref. \onlinecite{XinLi} with a simplified model Hamiltonian. In our calculation, we use DFT to get the full Hamiltonian which can describe the system more accurately.

The remainder of this paper is organized as follows: In Sec. II we describe the method used for studying the conductance in detail. In this part, the Green's function formalism and  B$\ddot{\textup{u}}$ttiker probes are reviewed. Then, we discuss the results in Sec. III. We first study the phase-coherent conductance of the four strands both with and without the backbone and find that the phase-coherent conductance is many orders of magnitude smaller than the experimental values of the conductance. We then include decoherence in these strands to understand how decoherence changes the conductance. We show that by using two different decoherence rates one can provide reasonable agreement with experiments. Finally, we make a brief conclusion in Sec. IV.

\section{Model and Method}

\begin{figure}
 \centering
    \subfigure[ ]
    {
        \includegraphics[width=2.0in]{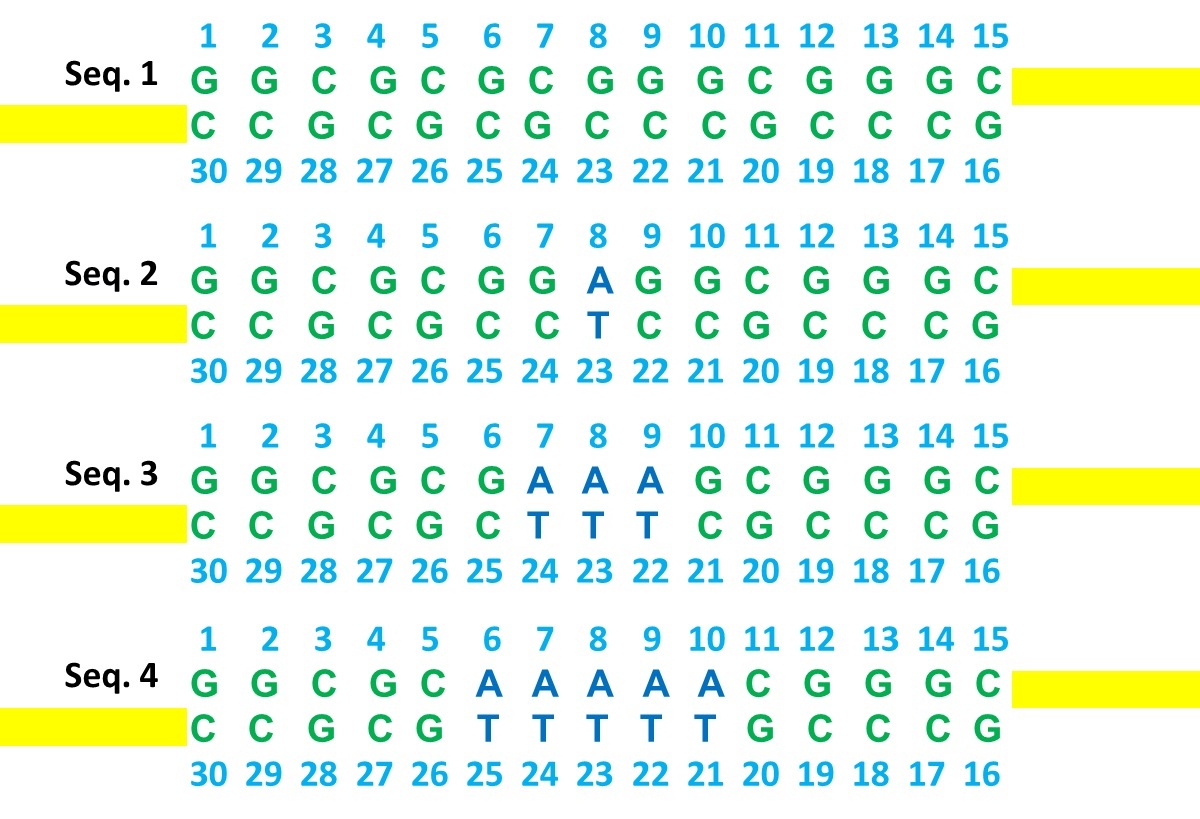}
        \label{Sequences}
    }
\\
  \subfigure[ ]
    {
        \includegraphics[width=2.0in]{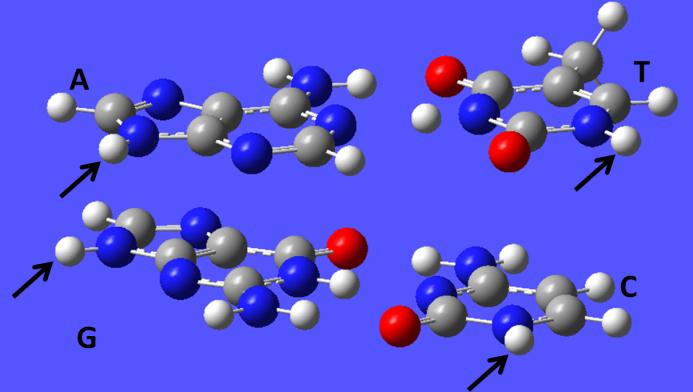}
        \label{Nobackbone}
    }
    \caption{(Color online) (a) Sketch of the four DNA sequences: Seq. 1, Seq. 2, Seq. 3 and Seq. 4, where each base is numbered. Yellow regions indicate the ends connecting to the contacts. (b) Atomic structures of the hydrogen atoms terminated DNA bases. Arrows indicate the hydrogen atoms that replace the backbones.}
    \label{fig1}
\end{figure}

We consider four different DNA strands, in which the five base pairs in the center part are changed from $G:C$ to $A:T$ [Fig.  \ref{Sequences}]. The choice of our model system is motivated by the experiment in Ref. \onlinecite{Mahapatro}, which provides data for the room temperature conductance of these four different strands under similar conditions, where the strands are connected to electrical contacts via thiol groups. The four strands we study consist of the following 15-base pair double-stranded DNA molecules: $\textup{(a)}~GGCGCGCGGGCGGGC;~\textup{(b)}~GGCGCGGAGGCG\\GGC;~\textup{(c)} ~GGCGCGAAAGCGGGC;~\textup{(d)}~GGCGCAAA\\AACGGGC$ [Fig. \ref{Sequences}]. Conductances of these strands were measured in  Ref. \onlinecite{Mahapatro} under dry condition. The polycation spermidine was used to stabilize the double-stranded DNA molecules. In these systems, both the base pairing and base stacking of B-DNA is expected to remain unaffected. \cite{Gosule2, DengH}

Our method involves the following five steps. First, we obtain the atomic coordinates for the double-stranded B-DNA using the Nucleic Acid Builder (NAB) software package. \cite{NAB} To study the effect of the backbone, we consider two types of strands - the native ones with the backbones and the ones whose backbones are replaced with the hydrogen atoms. The sketch of the latter case is shown in Fig. \ref{Nobackbone}, where the arrows indicate the hydrogen atoms that replace the backbone. Second, we use Gaussian 09 \cite{Frisch} to obtain the Hamiltonian and overlap matrices, $H_0$ and $S_0$, where the B3LYP functional and the  6-31G basis set \cite{Becke, CLee} are adopted. We then transform the system to an orthogonal basis using L$\ddot{\textup{o}}$ewdin transformation, \cite{POLowdin, Mehreztrans}
\begin{equation}\label{Hamiltonian1}
H_1=S_0^{-\frac{1}{2}} H_0 S_0^{-\frac{1}{2}}
\end{equation}
\begin{equation}\label{Hamiltonian}
H=U^{\dagger} H_{1} U
\end{equation}
In Eq. $(\ref{Hamiltonian})$, $U$ is a block diagonal matrix. To obtain $U$, we first diagonalize every diagonal sub-block of $H_1$ and then arrange the eigenvectors in the order of DNA bases. While the diagonal blocks of $H_0$ are full matrices, the diagonal blocks of $H$ are diagonal matrices and the dimension of the diagonal block matrices is equal to the number of orbitals used to represent that base. Physically, the diagonal blocks of $H$ correspond to the localized energy levels of each DNA base and off-diagonal blocks correspond to interactions between different bases. If we look at the values of the off-diagonal elements, we find that the interaction corresponding to energy levels at the two nearby bases is large while that corresponding to two far away bases is relatively small.

The transmission through the molecule is computed using the Green's function approach, \cite{Datta} with the B$\ddot{\textup{u}}$ttiker probes to account for the decoherence. We calculate the retarded Green's function, which is defined by,
\begin{equation}\label{Green}
[E-(H+\Sigma_{L}+\Sigma_{R}+\Sigma_{B} )] G^{r}=I
\end{equation}
where $H$ is the Hamiltonian shown in Eq. $(\ref{Hamiltonian})$. The self-energy of the left (right) contact  $\Sigma_{L(R)}$ represents the coupling of the DNA to left (right) contact through which charge enters and leaves the DNA. The self-energy due to the phase-breaking B$\ddot{\textup{u}}$ttiker probes  is $\Sigma_{B}$.

The self-energy due to the contacts is the third step of our calculations. The correct expressions for the self-energy due to the contacts are $\Sigma_{L}=T_{DL} g_{L} T_{LD}$ and $\Sigma_{R}=T_{DR} g_{R} T_{RD}$. Here $g_{L}$ and $g_{R}$ are the surface Green's functions of the contacts, $T_{LD}$ and $T_{RD}$ are the coupling between the left and right contacts and the device, with $T_{DL}=T_{LD}^{\dagger}$ and $T_{DR}=T_{RD}^{\dagger}$. Computing these self-energies for realistic contacts is challenging. To obtain an accurate answer, the inclusion of contact surface atoms with optimized structure and their influence on the device molecule are required though some progress has been made recently in this area. \cite{Lambert} In this paper, we neglect the real part of the self-energy and further set  $\Sigma_{L(R)}=-i\Gamma_{L(R)}/2$, where $\Gamma_{L(R)}$  is treated as an energy independent parameter. Mathematically, the coupling matrix is diagonal with the non-zero elements representing the coupling strength. This approximation has been adopted by other researchers in DNA transport,  \cite{Cuniberti, Zilly, MHLee}  especially at small biases as long as the value of $\Gamma_{L(R)}$ is close to experiments. We verify this approximation by trying different $\Gamma_{L(R)}$ values.   

\begin{figure}[h]
\includegraphics[width=3.5in]{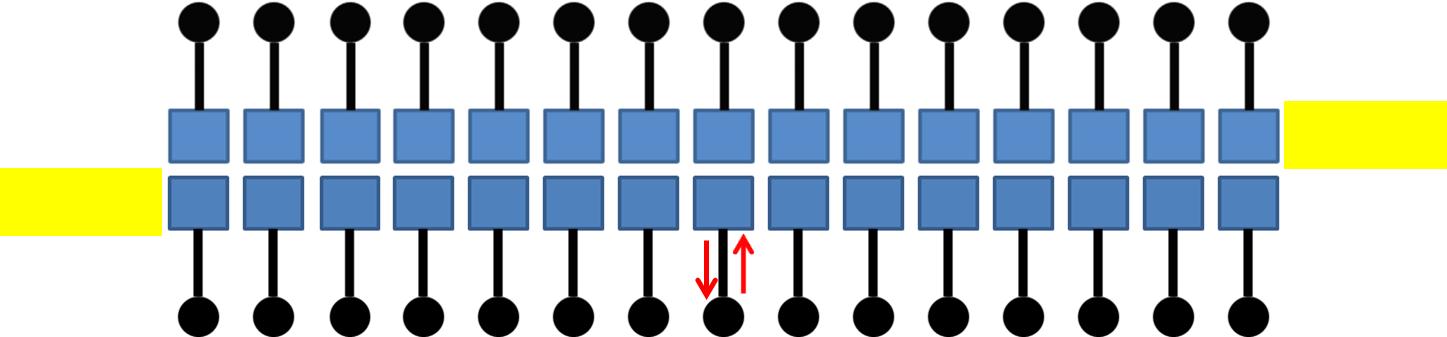}
\caption{(Color online) Sketch of the device with B$\ddot{\textup{u}}$ttiker probes. Yellow regions indicate the ends connecting to the contacts. Blue regions indicate DNA bases and black dots are B$\ddot{\textup{u}}$ttiker probes. Red arrows describe the behavior of electron at B$\ddot{\textup{u}}$ttiker probes: first extracted from the device and then re-injected back.}
\label{BProbes}
\end{figure}

The fourth step involves the inclusion of B$\ddot{\textup{u}}$ttiker probes. B$\ddot{\textup{u}}$ttiker probes are fictitious probes which extract electrons from the device and re-inject them after phase breaking, as illustrated in Fig. \ref{BProbes}. The net current at each B$\ddot{\textup{u}}$ttiker probe is zero. Similar to the effect of the left and right contacts, the effect of B$\ddot{\textup{u}}$ttiker probes is also included as a self-energy, $\Sigma_B$, as shown in Eq. $(\ref{Green})$. Specifically, we use $\Sigma_i$ to represent the decoherence at the $i$th probe, $\Sigma_B=\sum\limits_i{\Sigma_i}$. In our model, $\Sigma_i$ is controlled by an energy-independent coupling strength between the probe and the coherent system $\Gamma_{i}$, $\Sigma_{i}=-i\Gamma_{i}/2$. In the calculations, we attach the  B$\ddot{\textup{u}}$ttiker probes to the energy levels of the bases that correspond to the diagonal elements of $H$. The strength of coupling of electrons from the DNA to the  B$\ddot{\textup{u}}$ttiker probe is assumed to depend only on whether it is a $G:C$ or $A:T$ base pair. Note that at first we include the  B$\ddot{\textup{u}}$ttiker probes only in the $G:C$ base pairs as $A:T$ base pairs are tunneling barriers. \cite{SSairam} We find that only including decoherence on $G:C$ is not enough to explain the experiment. The decoherence on $A:T$ also plays a role in conduction, indicating that $A:T$ is not a coherent static barrier.

The fifth step uses the Green's function approach to calculate the conductance for various coupling strengths of the  B$\ddot{\textup{u}}$ttiker probes. The structure of the DNA is assumed to be static in the conventional B form throughout the calculation. The effect of fluctuations in the lattice and environment is included only via the Buttiker probes. In the following part of this section, we summarize the steps of obtaining the effective transmission and conductance after including the effect of B$\ddot{\textup{u}}$ttiker probes using the D'Amato-Pastawski model. \cite{Amato}

Assuming that the total number of DNA bases is $N$ (here, $N=30$), the number of B$\ddot{\textup{u}}$ttiker probes is $N_b=N-2$. At the low-bias region, the current at the $i$ th probe is,

\begin{equation}\label{Currenti}
I_{i}=\frac{2q}{h}\sum_{j=1}^{N} T_{ij} \lbrack \mu_{i}-\mu_{j} \rbrack ,  i=1,2,\cdots,N 
\end{equation}
	
The factor $2$ is used to account for the spin degeneracy. In the above equation, $T_{ij}$ is the transmission between the $i$th and $j$th probes calculated by $T_{ij}=\Gamma_{i} G^{r} \Gamma_{j} G^{a}$, where $G^a$ is the advanced Green's function, $G^a = (G^{r})^{\dagger}$. Because the transmission coefficients are reciprocal, we have $T_{ij}=T_{ji}$. \cite{Datta}

Using the condition that the net current is zero at each B$\ddot{\textup{u}}$ttiker probe, Eq. $(\ref{Currenti})$ gives us $N_b$ independent formulas, from which we can express the chemical potential of the $i$th B$\ddot{\textup{u}}$ttiker probe, $\mu_i$,
\begin{equation}\label{Mui}
\mu_{i}-\mu_{L}=\lbrack \sum_{j=1}^{Nb} W_{ij}^{-1} T_{jR}\rbrack(\mu_{R}-\mu_{L}), i=1,2,\cdots,N_b
\end{equation}

In the above equation, $W^{-1}$ is the inverse matrix of $W$, whose elements are given by, \cite{Nozaki} $W_{ij}=\lbrack(1-R_{ii} ) \delta_{ij}-T_{ij} (1-\delta_{ij})\rbrack$, where $R_{ii}$ is the reflection probability at probe $i$, which is given by $R_{ii}=1-\sum_{j \ne i}^{N}T_{ij}$.
The currents at the left contact $I_L$ and right contact $I_R$ are not zero. Because of the conservation of the electron number, $I_L+I_R=0$. Expressing the current at the left contact in terms of the difference between the chemical potentials at the two contacts $\mu_L$ and $\mu_R$,

\begin{equation}\label{CurrentL}
I_L=\frac{2q}{h} T_{eff} (\mu_L-\mu_R)	
\end{equation}

Comparing Eq. $(\ref{CurrentL})$ and the expression for the current at the left contact given by Eq. $(\ref{Currenti})$, we can write the effective transmission between left contact and right contact as,

\begin{equation}\label{Teff}
T_{eff}=T_{LR}+\sum_{i,j=1}^{N_b}T_{L,i} W_{ij}^{-1} T_{j,R} 
\end{equation}

The first term $T_{LR}$ is the coherent transmission from left contact to right contact while the second term describes the effect of decoherence. 
From Eq. $(\ref{CurrentL})$, the zero-bias conductance is approximately,

\begin{equation}\label{Cond}
G=\frac{2q^{2}}{h} T_{eff}
\end{equation}
$G_{0}=\frac{2q^2}{h} \approx 7.75 \times 10^{-5} \Omega^{-1}$ is the quantum conductance.

\section{Results and Discussion}
The sketch of the four 15-base pair DNA sequences that we model is shown in Fig. \ref{Sequences}. In the experiment, thiol groups are used to connect the 3' end of the DNA molecules to the metal contacts, because thiol groups are expected to provide a strong coupling between the DNA molecule and the metal contact. \cite{Mahapatro}  In this work, we use two broadening matrices, $\Gamma_L$ and $\Gamma_R$ to represent that coupling. We vary $\Gamma_L$ and $\Gamma_R$ between 50 and 100 meV, and find that there is no significant change in our results. For the results presented here, we set $\Gamma_L=\Gamma_R$ =100 meV.  

In this section, we analyze the effect of two factors in DNA charge transport using the Landauer-B$\ddot{\textup{u}}$ttiker formalism, namely, the backbone and the decoherence. In Section III A, we compare the transmission for the four strands with and without the backbones in the phase-coherent limit. We find that the backbone does not change the transmission qualitatively. However, it can affect the transport significantly at some energy points by interacting with the atoms in bases. In Section III B, we discuss the role of decoherence. We observe that the coherent transmission is too small to explain the experiment. We infer that the low value of conductance can be due to a number of reasons: poor coupling to contacts, position of Fermi energy or incorrect bandgap. We will show below that even for the best possible values of these parameters, the phase-coherent conductance is smaller than the experimental conductance. We will then show that by including appropriately chosen decoherence rates the calculated conductance values are comparable to the experiment. The strength of decoherence rate is difficult to compute $ab initio$. Here we have considered a variety of values to lend some insight into the ones which may be experimentally feasible. In the process of modeling decoherence, we have also found that it is possible to explain the experiments only if we assume that there is also decoherence in the $A:T$ barriers. That is, the $A:T$ barriers cannot be assumed to be static rectangular barriers.

\subsection{The Role of the Backbone }

To study the effect of backbone, we have calculated transmission for two cases - DNA strands with and without backbones in the phase-coherent limit. For the strands with backbone, we use the positively charged sodium ions $Na^{+}$ to  neutralize the backbone. To determine the geometry of the DNA strand after being neutralized, we first carry DFT calculation on a short strand containing five $G$ stacking bases with sodium ions placed nearby the phosphate group by setting the initial distances based on Ref. \onlinecite{HMehrez}. The atom coordinates for DNA molecule are fixed while the positions of $Na^{+}$ ions are relaxed. We approximate the optimized position of the sodium ion around the backbone of the middle $G$ in this short strand to be the positions of the thirty sodium ions in the long 15-base pair strand. We then carry out DFT calculation on the 15-base pair strand to obtain the self-consistent Hamiltonian and overlap matrices. For the strands without backbone, we simply delete the backbone and then terminate the base with hydrogen atoms using GaussView 5. \cite{Gaussview}

\begin{figure}[h]
          \includegraphics[width=3.5in]{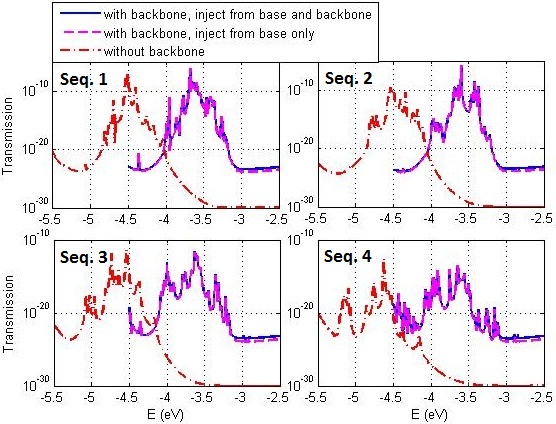}
    \caption{(Color online) Transmission vs energy for the four DNA strands with and without the backbones in the phase-coherent limit. The transmission for the strands with the backbones is calculated by two methods- charge is injected and extracted at the: (i) base and backbone (blue solid) and (ii) base only (magenta dash). Transmission for the strands without the backbone is shown by the red dash-dot curve. The transmission through the base only is slightly smaller than that through the base and backbone and the overall shapes of transmission with and without the backbone qualitatively seem to be similar but the two curves are shifted in the energy axis.}
\label{nodecoherence.}
    \label{Base_basebb}
\end{figure}

\begin{figure}[h]
          \includegraphics[width=3.5in]{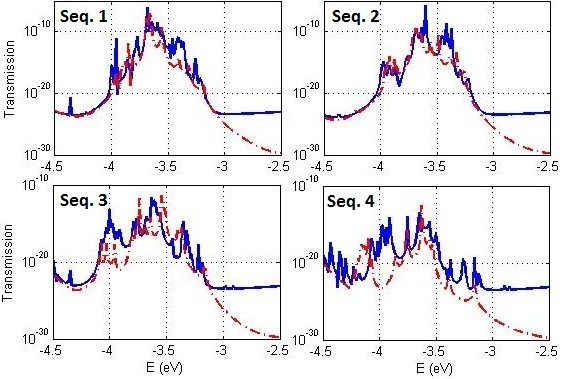}
    \caption{(Color online) Transmission vs energy for the four DNA strands with and without backbone in the phase-coherent limit. The transmission for the strands with backbone is shown in blue curves, while that for strands without backbone is shown in red with the energy axis shifted 0.85 eV for Seq. 1 and 2, and 1 eV for Seq. 3 and 4. The difference in magnitude of transmission for the strands with and without the backbones can be significant at some energy points.}
\label{nodecoherence.}
    \label{Base_basebb_shift}
\end{figure}

The transmission of the four DNA strands is calculated in the phase-coherent limit as shown in  Fig. \ref{Base_basebb}.  For the strands with backbone, we model the contacts via two approaches - charge is injected and extracted at the: (i) base and backbone (blue solid) and (ii) base only (magenta dash). Transmission for the strands without the backbone is shown by the red dash-dot curve of Fig. \ref{Base_basebb}. We note that the overall shapes of transmission with and without the backbone qualitatively seem to be similar but the two curves are shifted in the energy axis. To obtain a quantitative understanding, we shift the energy axis of the transmission for the strand without the backbone to the right-hand side by 0.85  eV for Seq. 1 and 2, and 1 eV for Seq. 3 and 4, as shown in Fig. \ref{Base_basebb_shift}. A close inspection of  Fig. \ref{Base_basebb_shift} reveals that the difference in magnitude of transmission for the strands with and without the backbones can be up to 5 orders for Seq. 1, 6 orders for Seq. 2, 7 orders for Seq. 3 and 8 orders for Seq. 4 at some energy points. 

\begin{figure}[h]
    \subfigure[ ]
    {
        \includegraphics[width=1.2in, height=2in]{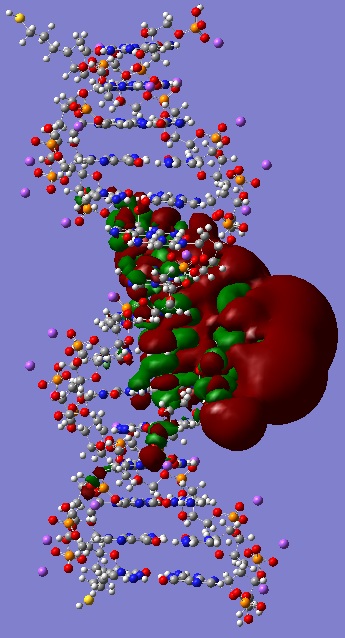}
        \label{Seq1_homo}
    }
    \subfigure[ ]
    {
        \includegraphics[width=1.2in, height=2in]{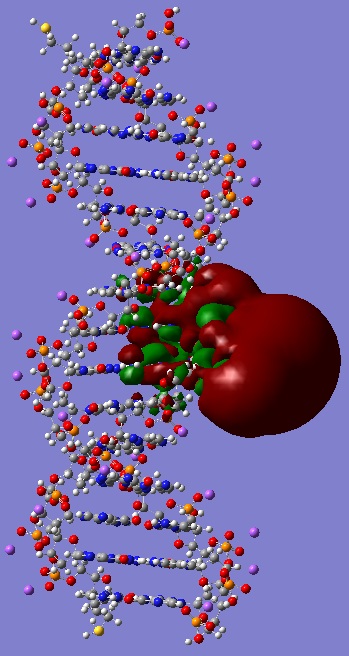}
        \label{Seq2_homo}
    }
    \subfigure[ ]
    {
        \includegraphics[width=1.2in, height=2in]{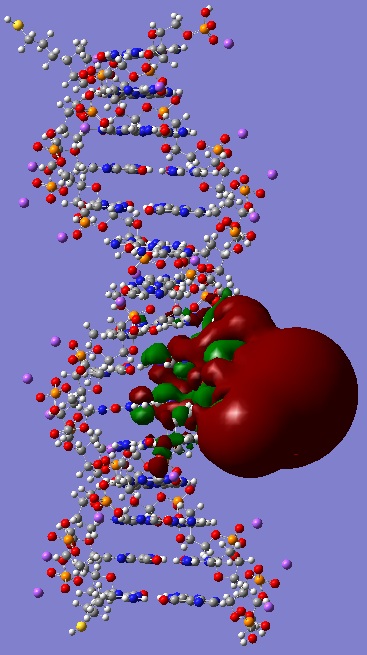}
        \label{Seq3_homo}
    }   
    \subfigure[ ]
    {
        \includegraphics[width=1.2in, height=2in]{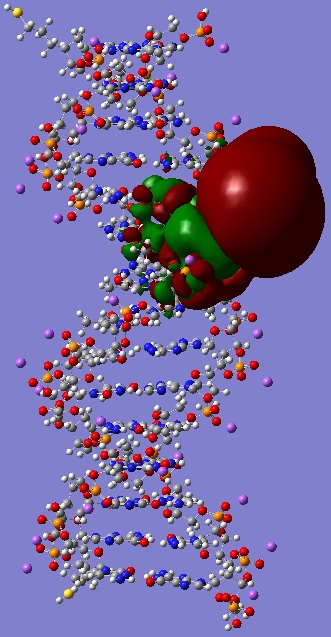}
        \label{Seq4_homo}
    }
    \caption{(Color online) Isosurfaces of HOMO orbitals for Seq. 1 (a),  Seq. 2 (b), Seq. 3 (c) and Seq. 4 (d). The red color represents the positive part of the wave function and the green color represents the negative part of the wave function. The HOMO orbital is distributed over both the bases and the backbone.}
    \label{Seq_homo}
\end{figure}

We study the effect of backbone further by plotting the HOMO orbitals for the four strands. HOMO orbitals for DNA strands with the backbones have been studied by a few groups. \cite{Barnett, NBarnett, MHLee} It has been shown that the spatial distribution of HOMO orbitals depends on a variety of conditions, such as the position of counterions relative to the phosphate groups \cite{Barnett} and the hydration levels. \cite{NBarnett} In addition, Ref. \onlinecite{MHLee} states that the HOMO orbitals can temporarily have a large weight on the backbones as a function of time. In our calculation, we find that the HOMO orbital is distributed over both the bases and the backbone, as shown in  Fig. \ref{Seq_homo}. We conclude from the above observations related to transmission ( Fig. \ref{Base_basebb_shift}) and wave function (Fig. \ref{Seq_homo}) that while most of the contribution to charge transport comes from the bases, the phosphate groups, sugar rings and sodium ions on the backbone alter the electronic structure sensitively by interacting with the bases. This leads to mismatches in the coherent transmission for strands with and without the backbone at some energy points while keeping the overall shape similar. We also note that the coherent transmission through the HOMO-LUMO gap also depends on the inclusion of the backbone. This finding indicates that when modeling the coherent charge transport in DNA molecules, one should take the effect of backbone into account. In the remainder of this paper, we will focus on the strands with the backbones, injecting and extracting from and into the strands via both backbone and base.    

\subsection{The Role of Decoherence}

Besides the interaction between the molecule and the metal contacts, the position of the Fermi level $E_f$ relative to the molecular energy levels is also important in understanding the flow of charge. $E_f$ depends sensitively on the surface condition and work function difference between two materials. Experimentally, it is very difficult to determine the position of $E_f$. Here, we explore the conductance when the Fermi level is in the HOMO vicinity. At the low-bias region, only electrons with energies nearby the Fermi level contribute to the transport. Thus, the transmission $T(E) \cong T(E_f)$. From DFT calculations, the HOMO levels of the 4 strands have been determined to be -3.18 eV, -2.84 eV, -2.91 eV, -2.85 eV, respectively. 

\begin{figure}[h]
\includegraphics[width=3.0in]{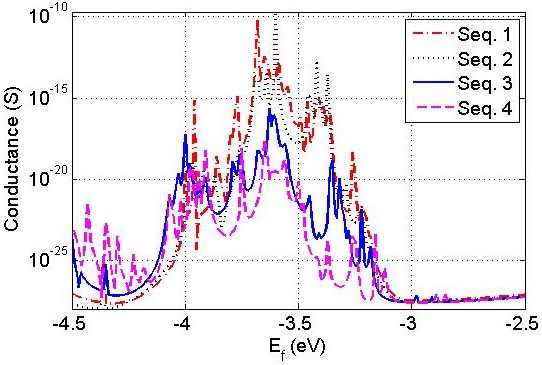}
\caption{(Color online) Conductance vs Fermi energy for the four DNA strands without any decoherence. The coherent conductance is orders of magnitude smaller than experiment irrespective of the locations of Fermi levels for all the four strands.}
\label{nodecoherence}
\end{figure}

The zero-bias conductance versus the position of Fermi energy is shown in Fig. \ref{nodecoherence}. Comparing the conductances of Seq. 2, Seq. 3 and Seq. 4, we find that the $A:T$ base pair plays the role of a barrier. This is consistent with the fact that in our calculation the ionization potential (IP) of an isolated $G:C$ base pair is 7.12 eV while that of the $A:T$ base pair is 7.68 eV. The similarity of the conductance for Seq. 1 and Seq. 2 can be attributed to the fact that Seq. 1 has a $G:C$ base pair in the sixth position and a $C:G$ base pair in the seventh position which can cause interstrand hopping, and results in lower transmission while in the Seq. 2 both the sixth and seventh positions are placed with $G:C$ base pairs. However, the presence of $A:T$ base pair in the eighth position cancels out this advantage. More importantly, in this coherent model, we find that the conductivity is much smaller than the experimental results. At the low-bias region, the experimental value \cite{Mahapatro} of the conductance for Seq. 1, Seq. 2, Seq. 3 and Seq. 4 is around $5\times 10 ^{-10}$ S, $2\times 10^{-10}$ S, $3.5 \times 10^{-11}$ S and $6 \times 10^{-12}$ S, respectively. However, in the coherent case of our calculation, the conductance is only about $10^{-24} S$ for Seq. 1 and $10^{-28}$ S for Seq. 2, Seq. 3 and Seq. 4  when the Fermi level is nearby the HOMO. For Seq. 1, the coherent result is about $10^{14}$ times smaller than the experiment; for Seq. 2, $10^{18}$ times smaller; for Seq. 3, $10^{17}$ times smaller and for Seq. 4, $10^{16}$ times smaller. 

The huge difference between the coherent conductance and experimental values indicates that there are other mechanisms that play a role in determining the conductance. We rule out the location of Fermi energy because irrespective of where the Fermi energy is in Fig. \ref{nodecoherence} (including well within the HOMO band), the conductance is significantly smaller than in experiments. We also rule out the nature of the coupling to contacts because we have verified that our results change by small amounts when the coupling to the contacts $\Gamma_L$ and $\Gamma_R$ is varied from 50 to 100 meV. Another reason for the larger experimental conductance could be the possibility of having a large number of strands, which is extremely unlikely in the break junction geometry of Ref. \onlinecite{Mahapatro}.

The mismatch in energy level between neighboring bases combined with the small interbase coupling leads to the smaller than experimental conductances in the phase coherent calculations. Decoherence due to interaction with the environment broadens these energy levels and can lead to a larger conductance. To account for decoherence, we use B$\ddot{\textup{u}}$ttiker probes as presented previously.

\begin{figure}[h]
    \subfigure[ ]
    {
        \includegraphics[width=3.0in]{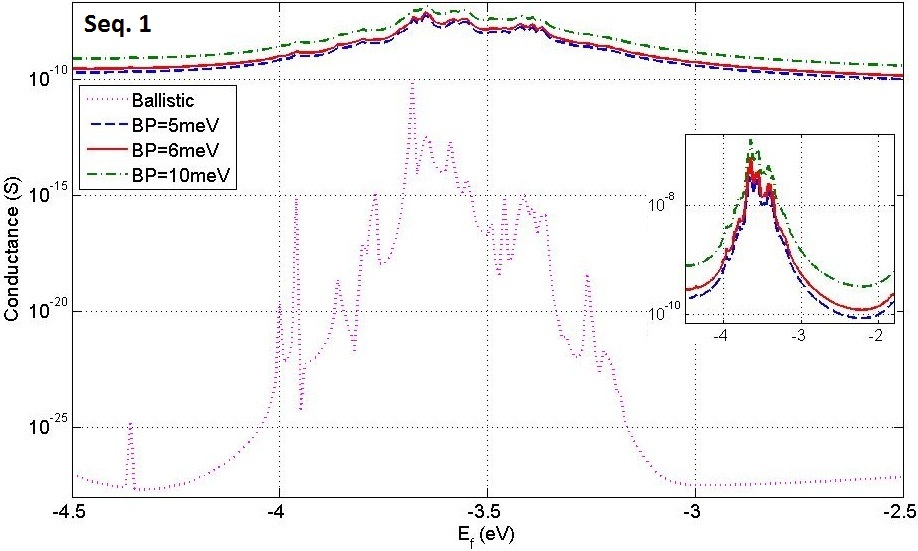}
        \label{BPonlyonGC_Seq1}
    }
    \\
    \subfigure[ ]
    {
        \includegraphics[width=3.0in]{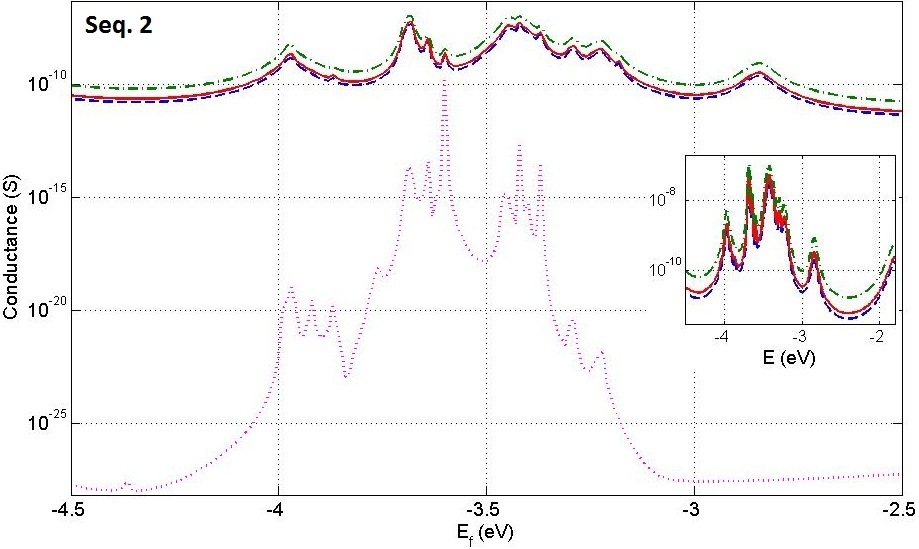}
        \label{BPonlyonGC_Seq2}
    }
   \\
    \subfigure[ ]
    {
        \includegraphics[width=3.0in]{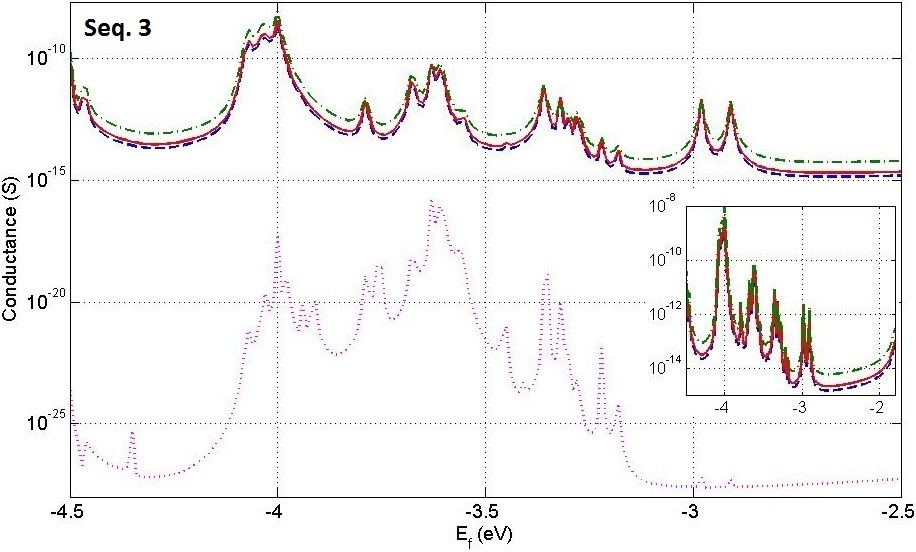}
        \label{BPonlyonGC_Seq3}
    }   
  \\
    \subfigure[ ]
    {
        \includegraphics[width=3.0in]{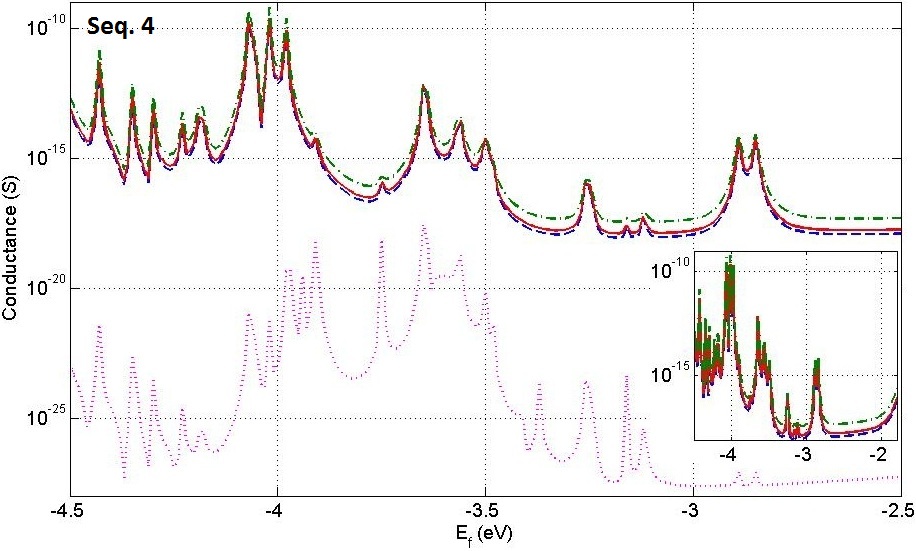}
        \label{BPonlyonGC_Seq4}
    }
    \caption{(Color online) The conductance versus Fermi energy with 5 meV, 6 meV and 10 meV decoherence only on $G : C$ base pairs for (a) Seq. 1, (b)  Seq. 2, (c) Seq. 3 and (d) Seq. 4. The conductance values for Seq. 3 and Seq. 4 around their HOMO levels are still too small to explain the experiment, suggesting the decoherence on $A : T$ barrier is also important.}
    \label{BPonlyonGC_Seq}
\end{figure}

As shown in Fig. \ref{nodecoherence}, for hole transport, $G:C$ base pair is preferred while $A:T$ base pair is considered as a barrier because the transmission decreases with increasing number of $A:T$ base pairs. As the transit time through a rigid barrier $A:T$ is small, we first include decoherence only on the $G:C$ base pairs with the coupling strength $\Gamma_i =$ 5 meV, 6 meV and 10 meV, as shown in Fig. \ref{BPonlyonGC_Seq}. Compared with the ballistic transport, one can find that after including B$\ddot{\textup{u}}$ttiker probes the conductance is smoother for the four strands. The small peaks in the ballistic conductance are smeared out due to the broadening of energy levels. For Seq. 1, we find that when we add 5 meV decoherence only on the $G:C$ base pairs, the conductance increases by around $10^{15}$ times around their HOMO levels, while for Seq.2, Seq. 3 and Seq. 4, the increase is $10^{18}$, $10^{16}$ and $10^{13}$ respectively. The tremendous enhancement in the conductance after including B$\ddot{\textup{u}}$ttiker probes suggests the crucial role of decoherence in charge transport through DNA molecules. 

In addition, we find that for strands containing $A:T$ base pairs, the same values of decoherence rates can enhance the conductance more effectively for Seq. 2 than Seq. 3 and Seq. 4. We observe that when the decoherence on $G:C$ base pairs is 5  meV, as the number of $A:T$ base pairs increases from 1 to 5, the increase in conductance decreases from around $10^{18}$ to $10^{13}$. This is because in Seq. 3 and Seq. 4, where the numbers of $A:T$ are 3 and 5, respectively, the wider barriers begins to play a larger role than in Seq. 2. We also find that when the decoherence on $G:C$ base pairs changes from 5 meV to 10 meV, the conductance increases around 4 times  for Seq. 1 and Seq. 2 around the HOMO levels. However, for Seq. 3 and Seq. 4, when the decoherence value on $G:C$ varies from 5 meV to 10 meV, the conductance only increases 1.5 times. More importantly, for Seq. 1 and Seq. 2, as the B$\ddot{\textup{u}}$ttiker probe coupling strength changes, we can fit the experimental conductance by setting the Fermi energy to be a particular value. For instance, when the decoherence on $G:C$ is 5 meV if we set the Fermi level to be -3.06 eV for Seq. 1 and -2.82 eV for Seq. 2, the computed conductance is comparable to that of experiment. However, for Seq. 3 and Seq. 4, where the wide $A:T$ barriers are present, irrespective of the strength of decoherence chosen (varying from 0 to 10 meV), the conductances are still too small when compared with the experiments. For example, for Seq. 3 the best conductance value is 10 times smaller than experiment and for Seq. 4, the best value is 100 times smaller. This indicates that the decoherence on $A:T$ barriers is also important, that is, the $A:T$ barrier is not a static barrier for hole transport. For this reason, we study the effect of decoherence on $A:T$ base pairs to get better agreement with the experiments.

\begin{figure}[h]
    \subfigure[ ]
    {
        \includegraphics[width=3.0in]{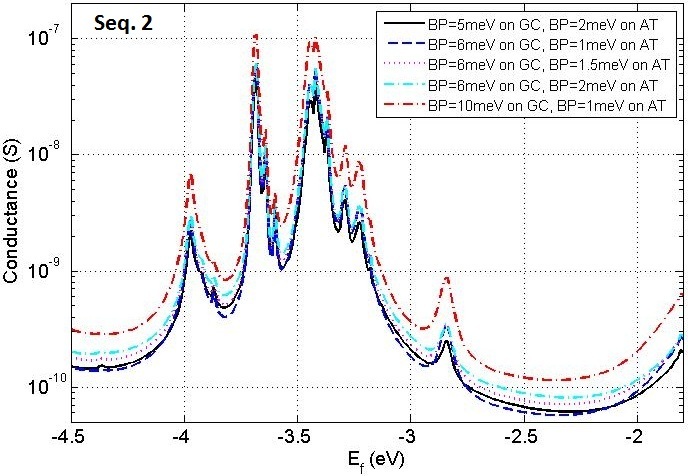}
        \label{BPonGCAT_Seq2}
    }
    \\
    \subfigure[ ]
    {
        \includegraphics[width=3.0in]{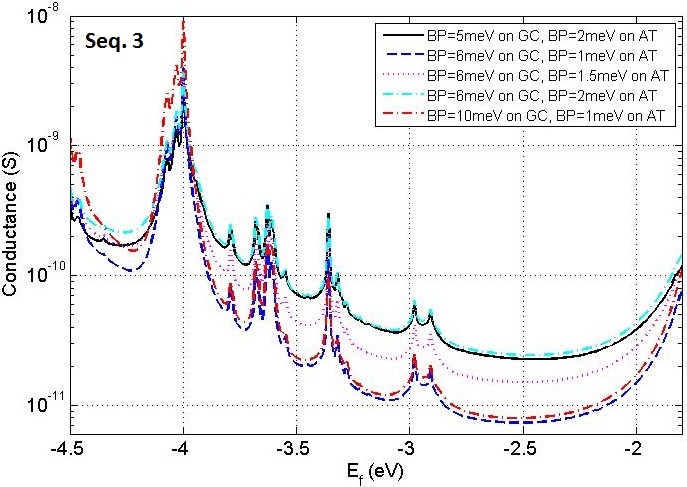}
        \label{BPonlyonGC_Seq3}
    }
   \\
    \subfigure[ ]
    {
        \includegraphics[width=3.0in]{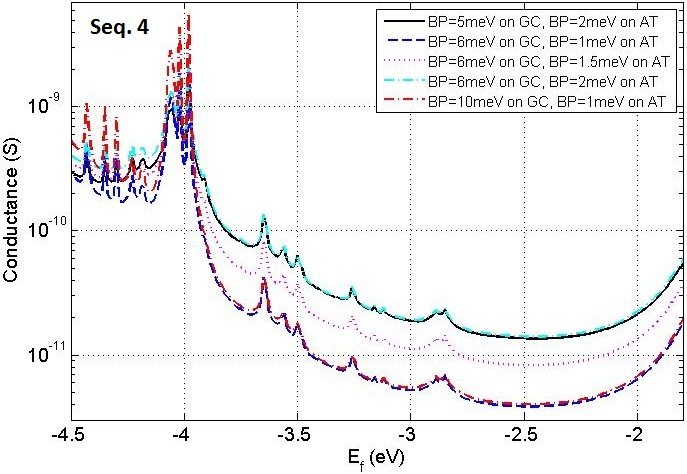}
        \label{BPonlyonGC_Seq4}
    }   
    \caption{(Color online) The conductance versus Fermi energy with different decoherence values on both $G : C$ and $A:T$ base pairs for (a) Seq. 2, (b)  Seq. 3 and (c) Seq. 4. The decoherence on $A : T$ base pairs can increase the conductance effectively, especially for Seq. 3 and Seq. 4.}
    \label{BPonGCAT}
\end{figure}

Fig. \ref{BPonGCAT} shows the conductance versus Fermi energy with different decoherence values on both $G:C$ and $A:T$ base pairs.  It is found that for Seq. 2, the decoherence on $G:C$ base pairs  is more important than that on $A:T$ base pair. For example, when the decoherence on $A:T$ base pair is fixed as 2 meV, the conductance increases 1.5 times with the decoherence on $G:C$ base pairs increasing from 5 meV to 6 meV. However, if the decoherence on $G:C$ base pairs is fixed as 6 meV, the coductance stays unchanged at the HOMO level when the decoherence on $A:T$ base pair is varied from 1 meV to 2 meV. In contrast, for Seq. 3 and Seq. 4, the decoherence on $A:T$ base pairs are much more important. Take Seq. 3 for example. If the decoherence on $A:T$ base pairs is fixed as 2 meV and the decoherence on $G:C$ is varied from 5 meV to 6 meV, the conductance only changes around 1.1 times. Similarly, if the decoherence on $A:T$ is fixed as 1 meV and the decoherence on $G:C$ is changed from 6 meV to 10 meV, the conductance also only changes around 1.1 times. However, if the decoherence on $G:C$ is fixed as 6 meV and the docoherence on $A:T$ is changed from 1 meV to 2 meV, the conductance increases around 2.8 times. 

\begin{figure}[h]
\includegraphics[width=3.0in]{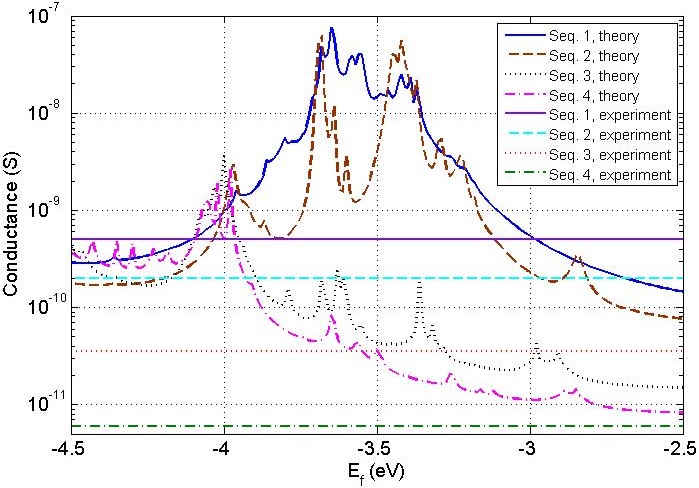}
\caption{(Color online) The conductance versus Fermi energy with $6~meV$ decoherence on $G:C$ and $1.5~meV$ decoherence on $A:T$. The conductance has been increased effectively. The conductance values for the 4 strands around their HOMO levels are quantatively comparable to the experimental results, demonstrating the importance of adding decoherence on both $G:C$ base pairs and $A:T$ barrier.}
\label{decoherenceGCAT}
\end{figure}

We find that for the four strands a decoherence strength of 6 meV on $G:C$ base pairs and 1.5 meV on $A:T$ base pairs give a relatively good agreement with experiments. The results are summarized in Fig. \ref{decoherenceGCAT}, where the experimental values of conductance from Ref. \onlinecite{Mahapatro} are also shown, with horizontal lines. We find that for all the four strands, the computed conductance lies in the ball park of the experimental values. That is, the qualitative trend of the conductance variation for the four sequences matches well with the experiments (Conductance decreases from Seq. 1 to Seq. 4). More importantly, in contrast to the phase coherent results in Fig. \ref{nodecoherence}, where the conductance values are orders of magnitude smaller than the experiments, the conductance values in  Fig. \ref{decoherenceGCAT}  are now comparable to the experiments.  This qualitative match holds well irrespective of the Fermi energy as can be seen in Fig. \ref{decoherenceGCAT}. Although for Seq. 4, the computed conductance is about 2 times larger than the experimental result around the HOMO level, we still think that the fitting is good because of the uncertain factors in the experiments.

These results indicate that the decoherence is important in experiments involving dry DNA, and this might be one of the reasons why the theory has difficulty in explaining experiments. 

Recently, there have been proposals to calculate the conductance of DNA by calculating the phase coherent conductance for many points (each of which provides a static set of distinct coordinates) along the trajectory of and MD simulation and averaging the coherently calculated conductance. \cite{Benjamin} These coordinates include the effect of time varying ions and water molecules, apart from thermally induced lattice vibrations. It would be interesting to compare results from the two methods for the case of dry DNA to see if one can match experiments on the linear response conductance and current-voltage characteristics.  To start with one could compare the average DOS obtained by the two methods. Also, work in Ref. \onlinecite{Benjamin} calculates the phase coherent conductance of DNA for many realizations of coordinates obtained from MD calculations of DNA in water. Such a process may give rise to an effective broadening of energy levels.

\section{Conclusion}
In summary, we have modeled the zero-bias conductance of dry DNA using a combination of density functional theory and the phenomenological B$\ddot{\textup{u}}$ttiker probes to account for decoherence. We first explore the effect of the backbone in charge transport by comparing the coherent transmission for strands with backbones and strands whose backbones have been deleted. We find that the DNA backbone can affect the coherent transmission significantly at some energy points. However, the overall shapes of the transmission are similar for the strands with backbones and strands without backbones. More interestingly, we find that the calculated conductance using phase coherent transport is orders of magnitude smaller than experiments, and that including the effect of decoherence on $G:C$ is crucial because it broadens the energy levels of DNA and significantly enhances the conductance. By comparing with experiments, we find that quite a large decoherence is required even on the $A:T$ base pairs, even though they behave as barriers in the energy ranges of importance. It is also worth noting that for the $G:C$ rich strands (strands which contain one $A:T$ base pair), the decoherence on $G:C$ base pairs is more important than that on $A:T$ base pairs, while for the strands contain three or five $A:T$ base pairs, the decoherence on $A:T$ plays a more significant role. By analyzing the four different DNA strands, it has been determined that the decoherence strength due to B$\ddot{\textup{u}}$ttiker probes is approximately 6 meV for $G:C$ base pairs and 1.5 meV for $A:T$ base pairs. While the phenomenological B$\ddot{\textup{u}}$ttiker probes is able to explain the experiments qualitatively at low biases, a more accurate description of vibronic coupling would be necessary to explain experiments at large biases. This is a significantly more difficult problem requiring knowledge of both the vibronic modes and the coupling strength in DNA, which is well beyond the extent of this work.

\section{Acknowledgement}
We acknowledge extensive discussions with Professor David Janes on experimental measurements of conductivtiy in dry DNA. This material is based upon work supported by the National Science Foundation under Grant No. 102781.

\end{document}